\documentclass{iau}

\usepackage{graphicx}
\usepackage[a4paper, lmargin=6cm, rmargin=1cm, tmargin=6.0cm, bmargin=2.5cm, footskip=2cm]{geometry}
\usepackage[colorlinks,linkcolor=blue, citecolor=blue]{hyperref}
\usepackage[authoryear]{natbib}

\title[Water delivery by hit-and-run collisions] %% give here short title %%
{Water delivery to dry protoplanets by hit-and-run collisions}

\author[C.~Burger, T.~I.~Maindl \& C.~Sch{\"a}fer]   %% give here short author list %%
{Christoph Burger$^{1,2}$, Thomas~I.~Maindl$^1$ \and Christoph~Sch{\"a}fer$^2$}
\affiliation{$^1$Department of Astrophysics, University of Vienna, \\T{\"u}rkenschanzstra{\ss}e 17, 1180 Vienna, Austria \\email: {\tt c.burger@univie.ac.at} \\[\affilskip]
$^2$Institut f{\"u}r Astronomie und Astrophysik, Eberhard Karls Universit{\"a}t T{\"u}bingen, \\Auf der Morgenstelle 10, 72076 T{\"u}bingen, Germany}

\pubyear{2020}
\volume{345}  %% insert here IAU Symposium No.
\jname{Origins: from the Protosun to the First Steps of Life}
\editors{Bruce G. Elmegreen, L. Viktor T\'oth, Manuel G\"udel, eds.}
\begin{document}

\maketitle

\begin{abstract}
Final water inventories of newly formed terrestrial planets are shaped by their collision history. 
A setting where volatiles are transported from beyond the snowline to habitable-zone planets suggests collisions of very dry with water-rich bodies.
By means of smooth particle hydrodynamics (SPH) simulations we study water delivery in scenarios where a dry target is hit by a water-rich projectile, focusing on hit-and-run encounters with two large surviving bodies, which probably comprise about half of all similar-sized collisions \citep{Genda2017_Hybrid_code_Ejection_of_iron-bearing_fragments}.

\keywords{Hydrodynamics, methods: numerical, planets and satellites: formation}
%% add here a maximum of 10 keywords, to be taken form the file <Keywords.txt>
\end{abstract}

\firstsection % if your document starts with a section,
              % remove some space above using this command.
\section{Methods}
We base our analysis in this work on existing collision simulations from \citet{Burger2018_Hit-and-run} and re-examine these data under the aspect of a dry target being hit by a water-rich projectile.
The scenarios comprise differentiated, self-gravitating, embryo-sized bodies, consisting of an iron core, a silicate mantle, and a water(ice) shell.
For details about the SPH code and the material model we refer the reader to \citet{Schaefer2016_miluphcuda} and \citet{Burger2018_Hit-and-run}.
Projectile and target have equal initial compositions, 25\% iron, 65\% silicates and a 10\% water mass fraction (wmf). This allows us to track individual contributions from the projectile/target, by tracking the respective SPH particles' origin.
For the results presented here we assume a dry target body by tracing only the projectile's water. The differences to a simulation with a really dry target are small \citep{Burger2018_Hit-and-run}, and should be negligible for the global quantities we are interested in.

\section{Results and discussion}

\begin{figure}
% \vspace*{-2.0 cm}
\begin{center}
 \includegraphics[width=5.4in]{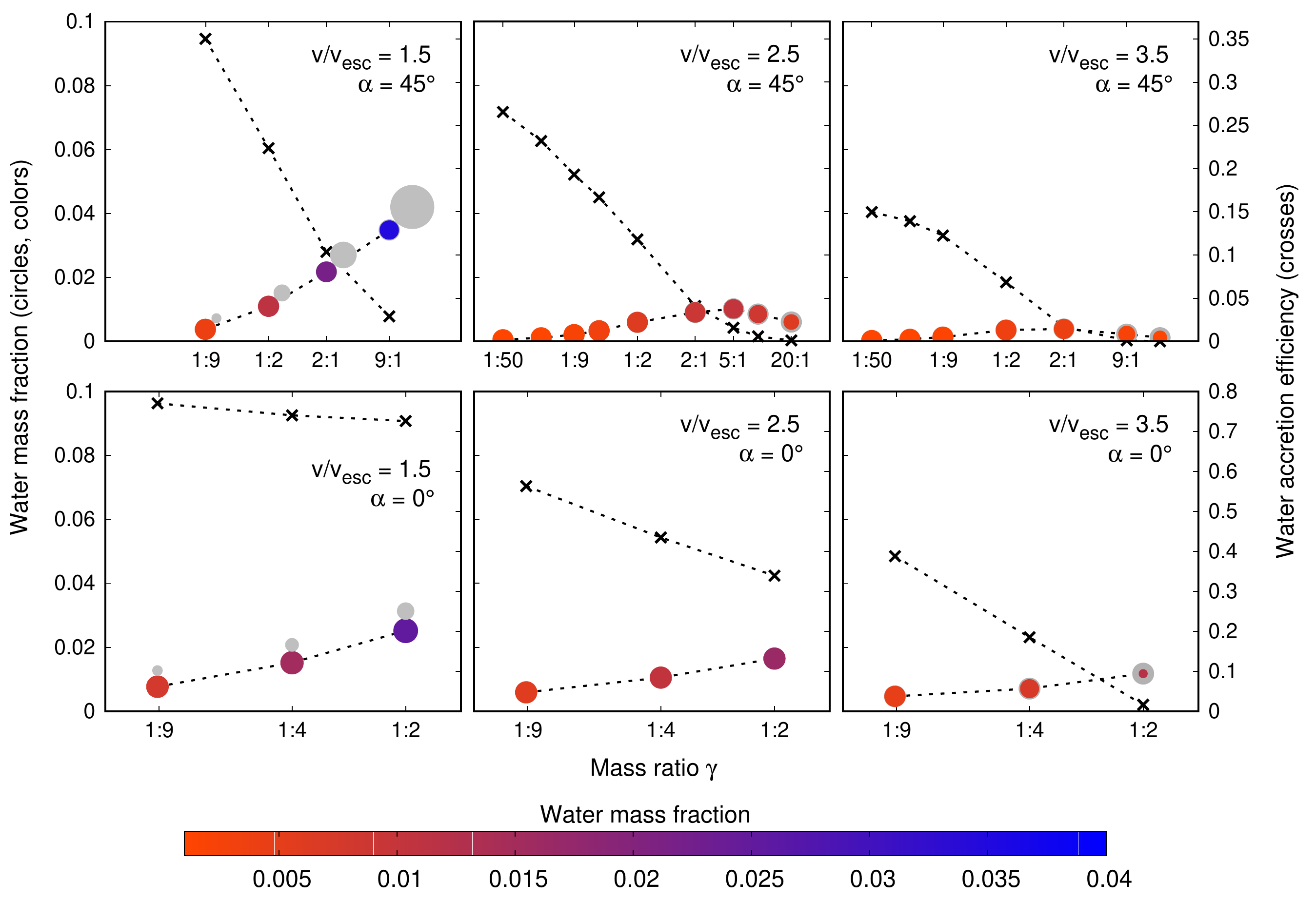}
% \vspace*{-1.0 cm}
 \caption{Dry targets being hit by water-rich projectiles (wmf = 0.1), for hit-and-run (top row) and head-on collisions (bottom row). Post-collision wmf are plotted on the left y-axes and color-coded. Water accretion efficiencies $\xi_\mathrm{w}$ are plotted on the right y-axes. Grey circles indicate either pre-collision sizes ($\propto \mathrm{mass}^{1/3}$), or the projectile body (impacting from the top).}
   \label{fig:dry_targets}
\end{center}
\end{figure}

Post-collision wmf of the initially dry target are plotted in Fig.~\ref{fig:dry_targets} for various parameter combinations (impact velocity $v/v_\mathrm{esc}$, impact angle $\alpha$, $\gamma = M_\mathrm{proj}/M_\mathrm{targ}$).
Typical hit-and-run collisions (top panels) are compared to head-on collisions (bottom panels). Along with that the water accretion efficiency is illustrated, here defined simply as $\xi_\mathrm{w} = M_\mathrm{w,tf} / M_\mathrm{w,p}$ with water mass on the target fragment and the projectile \citep[see also][]{Burger2017_hydrodynamic_scaling}. We abandon the usual notion of the projectile being the smaller body and switch to a one-body perspective, with an always dry target being hit by a (possibly even more massive) water-rich projectile.
In our low-velocity hit-and-run scenarios ($v/v_\mathrm{esc} = 1.5$, $\alpha = 45^\circ$ in Fig.~\ref{fig:dry_targets}) the target's post-collision wmf is strongly increasing with $\gamma$, and this trend continues even for projectiles considerably larger than the target. This behavior is qualitatively different for higher impact velocities ($v/v_\mathrm{esc} = \{2.5,3.5\}$, $\alpha = 45^\circ$), where the wmf accreted by the target increases with $\gamma$ and peaks at mass ratios of roughly 1:1, before decreasing again, despite the potentially huge water supply in larger projectiles.
The water accretion efficiency $\xi_\mathrm{w}$ -- the fraction of projectile water accreted by the target -- is a decreasing function of $\gamma$ for all scenarios in Fig.~\ref{fig:dry_targets}, and is generally high for small projectile bodies (small $\gamma$), but tends to be very low for larger impact velocities and/or mass ratios, indicating that only very little water is transferred to the target.

From the point of view of individual hit-and-run encounters the most water can be delivered to dry target bodies with either low impact velocities, or mass ratios approaching 1:1 for higher $v/v_\mathrm{esc}$.
However, planet formation is a chaotic process and planets grow from a stochastic history of successive collisions. Therefore it will be rather the (average) water accretion efficiency that determines how much water a growing planet can accrete from the limited amount of water-rich material scattered into its feeding zone. Our results show that this figure is considerably higher for smaller hit-and-run projectiles compared to larger ones, and yet significantly higher for head-on collisions, with water retention up to 80\%, compared to at most 35\% in our hit-and-run scenarios.
It is important to emphasize however, that this does not include the further fate of the (potentially still very water-rich) projectile, after a hit-and-run encounter, therefore it is crucial to consider and track \emph{both} hit-and-run fragments and their volatile inventories.

\end{document}